\documentclass[
   ,final            % use final for the camera ready runs
  ]
  {aipproc}
\usepackage{amssymb}
\layoutstyle{6x9}

\begin{document}

\title{Two-brane system in a vacuum bulk with a single equation of state}


\author{Juan L. P\'erez}{
  address={Departamento de F\'isica, DCI, Campus Le\'on, CP 37150,
    Universidad de Guanajuato,Le\'on, Guanajuato, M\'exico}
}

\author{Rub\'en Cordero}{
  address={Departamento de F\'\i sica, Escuela Superior de F\'\i sica y Matem\'aticas
del IPN, \\ Unidad Adolfo L\'opez Mateos, Edificio 9, 07738, M\'exico, Distrito
Federal, M\'exico}
}

\author{L. Arturo Ure\~na-L\'opez}{
  address={Departamento de F\'isica, DCI, Campus Le\'on, CP 37150,
    Universidad de Guanajuato,Le\'on, Guanajuato, M\'exico}
}

\begin{abstract}
We study the cosmology of a two-brane model in a five-dimensional
spacetime, where the extra spatial coordinate is compactifed on an
orbifold. Additionally, we consider the existence on each brane of
matter fields that evolve in time. Solving the Einstein equations in
a vacuum bulk, we can show how the matter fields in both branes
are connected and they do not evolve independently.
\end{abstract}

\maketitle

\section{Introduction}

The idea of brane worlds has been motivated in string theory, in which
our visible universe can be seen as a four dimensional "sheet" inmersed in
a spacetime of more spatial dimensions. Since only three of these
spatial dimensions are presently observable, one has to explain why
the others are hidden from detection \cite{cuerdas1,cuerdas2,cuerdas3}.
One such explanation is the so-called Kaluza-Klein (KK)
compactification, according to which the size of the extra dimensions is very small (see~\cite{k-k-1,k-k-2} for a review). In the Horava-Witten (HW)
solution~\cite{HW}, gauge fields of the Standard Model of Particle
Physics are confined in two 10-branes located at the end points of an
$S_{1}/Z_{2}$ orbifold. The 6 extra dimensions on the branes are
compactified in a very small scale close to the fundamental one, and
their effect on the dynamics is felt through "moduli" fields, i.e.,
through 5D scalar fields. A 5D realization of the HW theory and the corresponding brane-world
cosmology is given in \cite{300,301,302}. These solutions can be
thought of as effectively 5-dimensional, with an extra dimension that
can be large relative to the fundamental scale. They provide the basis
for the Arkani-Hamed-Dimopoulos-Dvali (ADD) \cite{ADD-1} and Randall-Sundrum (RS) \cite{randall,randall2}
brane models of 5-dimensional gravity. In the RS type 1 model, the
space-time necessarily contains two 3-branes, located, respectively,
at the fixed points $y = 0$, and $y=y_{c}$, where $y$ is the fifth
spatial dimension. The brane at $y = 0$ is usually called the hidden
(or Planck) brane, and the one at $y = y_{c}$ is called the visible
(or TeV) brane.

In the context of two-brane models with matter, it is natural to ask
if the evolution in time of the branes is related to one each other
(just as in the RS type 1 model, but in general for any
metric). Langlois~\cite{Binetruy:1999ut} has showed that
there exists a relationship between the energy density for the two
branes in the form of \emph{cosmological constrains}. The main goal of
this paper is to use the relations found by Langlois and analyze the
cosmology behind them, generalizing our previous results \cite{perez}.

\section{Mathematical background}
We begin with the most general 5-dimensional metric in which the
branes, located in $y=0$ ($y=0$-brane) and $y=y_{c}$ ($y=y_{c}$-brane), respectively, lie within an
homogeneous and isotropic subspace with curvature $k$ for each one,
and then
\begin{equation}
  \label{metrica}
  ds^{2} = - n^{2}(t,\left|y\right|) dt^{2} + a^{2}(t,\left|y\right|) g_{i j} dx^{i} dx^{j} +
  b^{2}(t,\left|y\right|) dy^{2} \, .	
\end{equation}
We imposed some symmetries upon the model: reflection,
$(x^{\mu},y) \rightarrow (x^{\mu},-y)$, and compactification,
$(x^{\mu},y) \rightarrow (x^{\mu},y+2my_{c}), \, m=1,2,\ldots $; and
we demand on each one of metric coefficients $a(t,\left|y\right|)$, $n(t,\left|y\right|)$ and
$b(t,\left|y\right|)$ to be subjected to the conditions \cite{wang}:
\begin{eqnarray}
  \label{condicion1}	
  \left[ F^\prime \right]_{0} &=& 2F^\prime |_{y=0+} \, , \\
  \label{condicion2}	
  \left[ F^\prime \right]_{c} &=& -2F^\prime |_{y=y_{c}-} \, , \\
  \label{condicion3}	
  F^{\prime \prime} &=& \frac{d^{2}
 F(t,\left| y \right|)}{d\left| y \right|^{2}} + \left[ F^\prime
  \right]_{0} \delta(y) + \left[ F^\prime \right]_{c} \delta(y-y_{c})
  \, .
\end{eqnarray}
In the above equations, the prime denotes derivate with respect to
$y$, the square brackets denotes the discontinuity in the first
derivative at the positions $y=0$ and
$y=y_{c}$. Eq.~(\ref{condicion3}) is obtained if we demand that
$\frac{d\left|y\right|}{dy} = 1$, and that $\frac{d^{2} \left| y
  \right|}{dy^{2}} = 2\delta(y) -2\delta(y-y_{c})$, for $y\in[0,y_{c}]$.
A subindex $0$ will be used for
quantities valued at $y=0$, whereas a subindex $c$ will be used for
quantities valued at $y=y_{c}$. Now, in order to obtain exact
dynamical solutions, we write the five-dimensional Einstein equations,
$\tilde{G}_{AB}+\Lambda_{5}g_{AB}=\kappa^{2}_{(5)}\tilde{T}_{AB}$, for the
metric~(\ref{metrica}),
\begin{eqnarray}
\label{einstein0}
\tilde{G}_{00} &=& 3 \frac{\dot{a}}{a} \left( \frac{\dot{a}}{a}+\frac{\dot{b}}{b}\right)-3\frac{n^{2}}{b^{2}}\left[\frac{a''}{a}+\frac{a'}{a}\left(\frac{a'}{a}-\frac{b'}{b}\right)\right]+3k\frac{n^2}{a^2},\\
\label{einstein1}
\tilde{G}_{ij}&=&\frac{a^{2}}{b^{2}}\delta_{ij}\left\{\frac{a'}{a}\left(\frac{a'}{a}+2\frac{n'}{n}\right)-\frac{b'}{b}\left(\frac{n'}{n}+2\frac{a'}{a}\right)+2\frac{a''}{a}+\frac{n''}{n}\right\} \nonumber \\
&+&\frac{a^{2}}{n^{2}}\delta_{ij}\left\{\frac{\dot{a}}{a}\left(-\frac{\dot{a}}{a}+2\frac{\dot{n}}{n}\right)-2\frac{\ddot{a}}{a}+\frac{\dot{b}}{b}\left(-2\frac{\dot{a}}{a}+\frac{\dot{n}}{n}\right)-\frac{\ddot{b}}{b}\right\}-k\delta_{ij}, \\
\label{einstein2}
\tilde{G}_{05}&=&3\left(\frac{\dot{a}}{a}\frac{n'}{n}+\frac{\dot{b}}{b}\frac{a'}{a}-\frac{\dot{a}'}{a}\right), \\
\label{einstein3}
\tilde{G}_{55}&=&3\frac{a'}{a}\left(\frac{a'}{a}+\frac{n'}{n}\right)-3\frac{b^{2}}{n^{2}}\left[\frac{\ddot{a}}{a}+\frac{\dot{a}}{a}\left(\frac{\dot{a}}{a}-\frac{\dot{n}}{n}\right)\right]-3k\frac{b^2}{a^2}.
\end{eqnarray}
We assume that the energy-momentum
tensor takes the form
\begin{equation}
  \label{tme}
  {\tilde{T}^{A}}_{B} =\frac{\delta(y)}{b_{0}}
  diag(-\rho_{0},p_{0},p_{0},p_{0},0) + \frac{\delta(y-y_{c})}{b_{c}}
  diag(-\rho_{c},p_{c},p_{c},p_{c},0) \, .
\end{equation}
Using the
Bianchi identity, $\nabla_{A} {\tilde{G}^{A}}_{B} =0 $, we obtain the
conservation equation for the energy density in the $y=0$-brane,
\begin{equation}
  \label{conservacion}
  \dot{\rho_{0}} + 3 \frac{\dot{a}_{0}}{a_{0}} (p_{0} + \rho_{0}) = 0
  \, .
\end{equation}

According to the Israel's junction conditions \cite{israel}, we need to describe
the presence of an energy density in terms of a discontinuity in the
metric across the origin in the extra spatial coordinate. So,
following Wang \cite{wang}, we obtain the metric coefficients that
satisfy the following boundary conditions:
\begin{equation}
  \frac{\left[a'\right]_{0}}{a_{0}b_{0}} =
  -\frac{\kappa^{2}_{(5)}}{3}\rho_{0} \, , \quad
  \frac{\left[n'\right]_{0}}{n_{0}b_{0}} =
  \frac{\kappa^{2}_{(5)}}{3} (3p_{0} + 2\rho_{0})=
  \frac{\kappa^{2}_{(5)}}{3} \rho_{0}(2+3\omega_{0}) \, . \label{salto2}
\end{equation}
\begin{equation}
  \frac{\left[a'\right]_{c}}{a_{c}b_{c}} =
  -\frac{\kappa^{2}_{(5)}}{3}\rho_{c} \, , \quad
  \frac{\left[n'\right]_{c}}{n_{c}b_{c}} = \frac{\kappa^{2}_{(5)}}{3}
  (3p_{c} + 2\rho_{c})=
  \frac{\kappa^{2}_{(5)}}{3} \rho_{c}(2+3\omega_{c}) \, . \label{salto4}
\end{equation}
Here, we have assumed a perfect fluid in both branes with $p_{0}=\omega_{0}\rho_{0}$ and $p_{c}=\omega_{c}\rho_{c}$.
We now proceed to solve the Einstein equations. Integrating Eq.~(\ref{einstein0}), we
obtain~\cite{Langlois},
\begin{equation}
\label{friedmann3}
\left(\frac{a'}{ab}\right)^2-\left(\frac{\dot{a}}{an}\right)^2 =ka^{-2}-\frac{\Lambda_{5}}{6}+C_{DR}a^{-4},
\end{equation}
where $C_{DR}$ is the called \emph{dark radiation}. On the other
hand, Eq.~(\ref{einstein2}), can be written as
\begin{equation}
  \label{0-5}
  \frac{\dot{b}}{b}=\frac{n}{a'}\left[\frac{\dot{a}}{n}\right]'.
\end{equation}
Because of the orbifold symmetry, we are interested in the exact solution of Eqs.~(\ref{friedmann3}),
and~(\ref{0-5}) only in the $[0,y_{c}]$ interval, which we solve in the next
section.

\section{Exact solutions for a vacuum bulk}
Inspired in a previous work \cite{perez}, we
will find an expression relating the evolution in time for the Hubble
parameter in our brane universe, which corresponds to the brane
located at $y=y_{c}$, when the hidden brane at $y=0$ is dominated by a
single matter component.

If we take the ansatz
\begin{equation}
  \label{ansatz}
  \frac{\dot{a}}{n}=\lambda(t)a^{m/2},
\end{equation}
and substitute it in Eqs.~(\ref{friedmann3}), and (\ref{0-5}), we find
that
\begin{eqnarray}
  \label{ecuacionb}
  b &=& a^{m/2} \, , \\
  \label{completa}
  a' &=& \epsilon a^{m/2} \left[ \lambda^{2} a^{m} + k -
    \frac{\Lambda_{5}}{6} a^{2} + C_{DR} a^{-2} \right]^{1/2} \, .
\end{eqnarray}
When $m=0,2,-2$, the $\lambda^{2}$ term behaves as curvature,
cosmological constant, and dark radiation, respectively. The term,
$\epsilon=\pm 1$ is the sign of the square root in
Eq.~(\ref{completa}).  For simplicity, we consider in this paper a
vacuum bulk and a flat geometry in each brane,
$k=\Lambda_{5}=C_{DR}=0$. So, integrating (\ref{completa}) and using
the boundary conditions~(\ref{salto2}) and~(\ref{salto4}) with
aid of Eq.~(\ref{condicion1}), we obtain
\begin{equation}
  \label{solucion1}
  a(t,y) = a_{0}(t) \left[ 1 + (m-1) \frac{\kappa^{2}_{(5)}}{6}
    \rho_{0} b_{0} y \right]^{1/(1-m)} \, ,
    \end{equation}
\begin{equation}    	
  \label{solucion2}	
  n(t,y) = n_{0}(t) \left[ 1 + \left( \frac{m}{2} + 2 + 3 \omega_{0}
    \right) \frac{\kappa^{2}_{(5)}}{6} \rho_{0} b_{0} y \right] \left[
    1 + (m-1) \frac{\kappa^{2}_{(5)}}{6} \rho_{0} b_{0} y
  \right]^{m/(2-2m)} \, ,
  \end{equation}
  \begin{equation}
  \label{solucion3}
  b(t,y) = b_{0}(t) \left[ 1 + (m-1) \frac{\kappa^{2}_{(5)}}{6}
    \rho_{0} b_{0} y \right]^{m/(2-2m)} \, .
\end{equation}
Notice that for $m=0$ we recover the linear solutions founded by
Langlois~\cite{Binetruy:1999ut}, and for $m=2$ and $\omega_{0}=-1$, we
find a conformal RS metric \footnote{Even though it is similar to the metric for a
Randall-Sundrum cosmology, we need non-constant values of $a_{0}$,
$n_{0}$ and $b_{0}$. Note in this case, from Eq.~(\ref{conexion0}), we also
have $\rho_{0}=-\rho_{c}$. To recover the RS solutions, we should keep
$\Lambda_{5}\neq 0$ and $\lambda=0$ in Eq.~(\ref{completa}).}. We omite here the case $m=1.$

So far, we have found a family of exact solutions which satisfy the 5D
Einstein equations in a vacuum bulk. Imposing on these solutions the
boundary conditions~(\ref{salto2}) and (\ref{salto4}), we obtain
\begin{eqnarray}
  \label{conexion0}
  \rho_{c} &=& -\rho_{0} \left[ 1 + (m-1) \frac{\kappa^{2}_{(5)}}{6}
    \rho_{0} b_{0} y_{c} \right]^{(m-2)/(2-2m)} \, , \\
  \label{conexion1}
  (m/2 + 2 + 3\omega_{c} ) &=& \frac{(\frac{m}{2} + 2 + 3\omega_{0})
    \left[ 1 +(m-1) \frac{\kappa^{2}_{(5)}}{6} \rho_{0} b_{0} y_{c}
    \right]}{\left[ 1 + (\frac{m}{2} + 2 + 3\omega_{0})
      \frac{\kappa^{2}_{(5)}}{6} \rho_{0} b_{0} y_{c} \right]} \, .
\end{eqnarray}

We use in Eq.~(\ref{conexion1}) that $n_{c}=1$ and $n_{0}=n_{0}(t)$, and
demand that the metric is of the FRW form in the $y=y_{c}$-brane. From
Eq.~(\ref{friedmann3}), and considering the particular case $k =
\Lambda_{5} = C_{DR}=0$, we find that
\begin{equation}
  \label{Friedmann3a}
  \frac{\dot{a}}{an}=\epsilon\ \frac{a'}{ab} \, .
\end{equation}
Taking into account the boundary conditions (\ref{salto2}) and~
(\ref{salto4}), the solutions~(\ref{solucion2}), together with
$n_{c}=1$ and also Eq.~(\ref{conexion0}), we obtain that
\begin{eqnarray}
  \label{friedmann3b}
  \label{hubble0}
  H_{0}=\frac{\dot{a}_{0}}{a_{0}}&=&-\epsilon\
  \frac{\kappa^{2}_{(5)}}{6}n_{0}\rho_{0}=-\epsilon\
  \frac{\kappa^{2}_{(5)}}{6}\rho_{0}\frac{\left[1+(m-1)\frac{\kappa^{2}_{(5)}}{6}\rho_{0}b_{0}y_{c}\right]^{-m/(2-2m)}}{\left[1+\left(\frac{m}{2}+2+3\omega_{0}\right)\frac{\kappa^{2}_{(5)}}{6}\rho_{0}b_{0}y_{c}\right]}
  \, , \\
\label{hubbleC}
H_{c}=\frac{\dot{a}_{c}}{a_{c}}&=&+\epsilon\
\frac{\kappa^{2}_{(5)}}{6}n_{c}\rho_{c}=-\epsilon\
\frac{\kappa^{2}_{(5)}}{6}\rho_{0}\left[1+(m-1)\frac{\kappa^{2}_{(5)}}{6}\rho_{0}b_{0}y_{c}\right]^{(m-2)/(2-2m)}
\,.
\end{eqnarray}
Eq.~(\ref{hubble0}), together with Eq.~(\ref{conservacion}), gives a
solution for $\rho_{0}(t)$ when $\omega_{0}$ is a constant, and from
it we obtain the evolution in time for $H_{c}(t)$. The sign $\epsilon$
is chosen such that we obtain an expanding universe within the
$y=y_{c}$-brane.

\begin{figure}[h]
	\centering
		\includegraphics[width=0.50\textwidth]{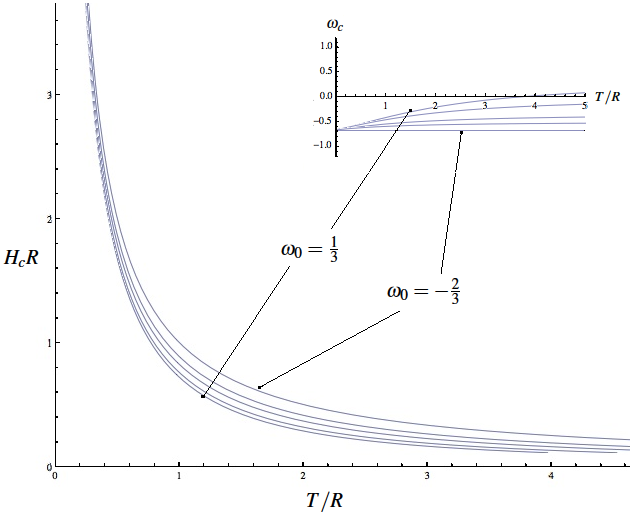}\  \  \ \includegraphics[width=0.50\textwidth]{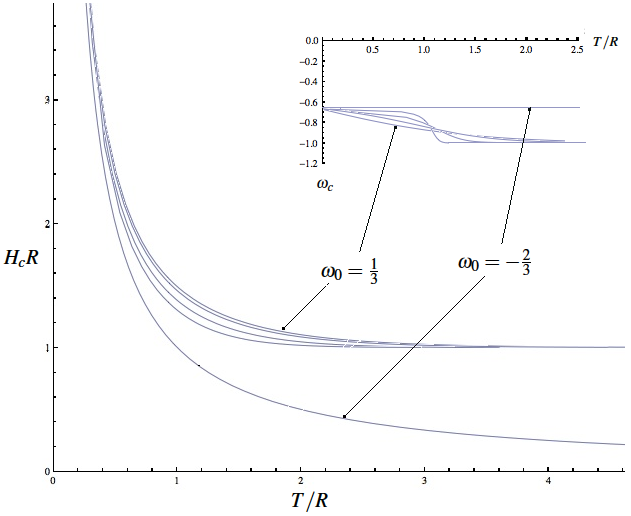}
	\caption{Evolution of $H_{c}R$ as a function of $T/R$ when $m=0$ and $\omega_{0}\in[-2/3,1/3]$, see Eqs.~(\ref{casom=0}). (Left) The case $0<X<1$ and $\epsilon=-1$; note that $H_{c}R$ approaches zero when $T/R>>1$. (Right) The case $X>1$ and $\epsilon=+1$; here $H_{c}R$ approaches the unity when $T/R>>1$, except  for $\omega_{0}=-2/3$, which is shown separately. The insets in both figures show, from Eq.~(\ref{conexion1}), the evolution of $\omega_{c}$ as a function of $T/R$ in each case.}
	\label{fig:E+}
\end{figure}

As an example, let us consider the case $m=0$ and $\omega_{0}=$ const., i. e. a bulk metric that is linear in $y$, and the $y=0$-brane is dominated by a single component. From Eqs.~(\ref{conservacion}), (\ref{hubble0}) and (\ref{hubbleC}), we obtain       
\begin{equation}\label{casom=0}
1-X^{-1}+(2+3\omega_{0})lnX=\epsilon\ (3+3\omega_{0})T/R \qquad  and \qquad  H_{c}R=-\epsilon\ \frac{X}{1-X},
\end{equation}
where $T=t-t_{*}$ , $X=\rho_{0}/\rho_{0*}$ , $\rho_{0*}=\rho_{0}(t_{*})$ , $R=b_{0}y_{c}$ is the radius of compactification, and $\omega_{0}\neq-1$. The constant $t_{*}$ is an epoch in which $\rho_{0*}^{-1}=\kappa^{2}_{(5)}b_{0}y_{c}/6$, i. e. proportional to $R$. There exist two cases that guarantee positive solutions for $H_{c}$ in the $y=y_{c}$-brane, namely $0<X<1$ when $\epsilon=-1$ and $X>1$ when $\epsilon=+1$, which are shown in Fig. \ref{fig:E+}.

\section{Conclusions}
We have showed, that in a two-brane system in a 5-D background space-time, there exists a relationship
between the cosmologies on the two branes. As in a RS setup, we considered a $S_{1}/Z_{2}$
compactification, and that a FRW metric is recovered in the
$y=y_{c}$-brane. We were able to generalize the results founded by Langlois et
al in\cite{Binetruy:1999ut} and our results are in concordance with a RS type cosmology. 

We found an expression relating the energy density in each brane
as well as the relation between the equations of state of the branes. Finally, we were able
to write the evolution in time for $H_{c}$, in the $y=y_{c}$-brane,
when the $y=0$-brane is dominated by a single matter component with a
constant equation of state $\omega_{0}$. These results may have an interesting interpretation for the cases of dark matter and dark energy in brane universes; we are currently exploring their possibilities, and expect to publish the results elsewhere.  

\begin{theacknowledgments}
We are grateful to the Divisi\'on de Gravitaci\'on y F\'isica
Matem\'atica (DGFM) for the opportunity to present this work at the
IX Taller of the DGFM, Colima 2011. We acknowledge partial support by
SNI-M\'exico, CONACyT research grant J1-60621-I, COFAA-IPN and SIP-IPN
grant 20121648. Besides, this work was partially supported by PIFI, PROMEP,
DAIP, and by CONACyT M\'exico under grants 167335, and the Instituto
Avanzado de Cosmologia (IAC) collaboration.
\end{theacknowledgments}

\bibliographystyle{aipproc}

\bibliography{BIBLIOGRAFIA-BRANAS}

\end{document}